\newcommand{\diracslash}[1]{#1\llap{/\kern2pt}}
\newcommand{\be}{\begin{equation}}
\newcommand{\ee}{\end{equation}}
\newcommand{\bea}{\begin{eqnarray}}\index{\footnote{}}
\newcommand{\eea}{\end{eqnarray}}
\newcommand{\ba}[1]{\begin{array}{#1}}
\newcommand{\ea}{\end{array}}

\documentclass[12pt]{iopart}

\usepackage{graphicx}
\usepackage{iopams}
\usepackage{setstack}
\usepackage{revsymb}
\begin{document}
\title{A model study on a pair of trapped particles interacting with an arbitrary effective range} 
\author{ Partha Goswami$^1$ and Bimalendu Deb$^{1,2}$ }
\address{$^1$ Department of Materials Science, $^2$ Raman Centre for Atomic, Molecular and Optical Sciences,
Indian Association for the Cultivation of Science,
 Jadavpur, Kolkata 700032. INDIA.}
\begin{abstract}
We study the effects of the effective range of interaction on the eigenvalues and eigenstates of two particles confined in a three-dimensional (3D) isotropic as well as one- or quasi-one dimensional harmonic (1D) traps. For this we employ model potentials which mimic finite-range $s$-wave interactions over a wide range of $s$-wave scattering length $a_s$ including the unitarity limits $a_s \rightarrow \pm\infty$. Our results show that when the range is larger than the 3D or 1D harmonic oscillator length scale, the eigenvalues and eigenstates are nearly similar to those of noninteracting two particles in the 3D or 1D trap, respectively. In case of 3D, we find that when the range goes to zero, the results of contact potential as derived by Busch {\it et al.} [Foundations of Physics, {\bf28}, 549 (1998)] are reproduced. However, in the case of 1D, such reproducibility does not occur as the range goes to zero. We have calculated the eigenvalues and eigenstates in 1D harmonic trap taking one-dimensional finite-range model potential. We have also calculated bound state properties of two particles confined in a highly anisotropic quasi-1D trap taking three-dimensional finite-range model potential, and examined whether these quasi-1D results approach towards 1D ones as the aspect ratio $\eta$ of the radial to axial frequency of the trap increases. We find that if the range is very small compared to the axial size of the trap, then one can reach 1D regime for $\eta \ge 10000$. However, for large range, one can nearly get 1D results for smaller values of $\eta$. This study will be important for exploration of two-body or many body physics of trapped ultracold atoms interacting with narrow Feshbach resonance for which the effective range can be large.
\end{abstract}
\maketitle

\section{Introduction}
Interacting systems in low dimensions have interesting or exotic features. For example, interacting electrons in two dimension exhibit fractional quantum Hall effect and high temperature superconductivity, interacting ultracold atoms in one dimension show Tonk-Girardeau regime
\cite{Tonks,Girardeau,Lieb,Paredes,Kinoshita}, confinement-induced resonances \cite{PRL:Olshanii:1998, PRL:Olshanii:2003, PRL:Moritz:2005}, metastable supercurrent states \cite{Kagan_2000}, etc. With the recent advancement of cooling and trapping of atoms by optical or magneto-optical forces, a new class of low dimensional systems, namely cold atoms in highly anisotropic harmonic traps, has become important. Interacting cold atoms in tightly confined or low dimensional traps have attracted a lot of research interests \cite{PRL:Olshanii:1998, PRL:Olshanii:2003, PRL:Moritz:2005, Petrov_01, Bolda_03, Idziaszek_05,Chin_11,Chevy_02,Vogt_12,Perin_13,Baskaran_89}. The interactions between neutral atoms are, in general, short-range unlike that in electrons where interactions are long-range due to Coulomb forces that may be modified due to lattice vibrations and confinement effects. Interactions in ultracold dilute atomic gases are usually described by a zero-range Fermi pseudo-potential \cite{Englert:1998} or a delta function potential expressed in terms of $s$-wave scattering length ($a_s$) only. The validity of zero-range potentials breaks down for narrow resonant interactions, dense atomic clouds and mixtures of different atomic gases in strong external electric fields. Finite-range effects become particularly important for narrow Feshbach resonances in ultracold atoms. It is therefore important to investigate the role of the range of interactions on interacting trapped atomic gases. For this we use finite-range model potentials recently proposed \cite{IJP:Deb:2015,Ijmp:Deb_2016} and shown to represent atom-atom interactions at low energy for arbitrary $a_s$ including unitarity-limited regime where $a_s$ diverges. Scattering length in cold atoms can be varied over a wide range by a magnetic Feshbach resonance \cite{Stoof_93,Kohler_06,Chin_10}. Near the unitarity regime, cold atoms become strongly interacting. Recently, it has been experimentally shown that the effective range of interactions between ultracold atoms near a magnetic Feshbach resonance can become quite large \cite{Harabati_99,Stoof_04,Pethick_05,Castin_06,Zoller_08,Li_12,Ohara_12,Thomas_12} leading to significant energy-dependence of the interactions. This calls for formulation of the problem beyond delta potential approach.

For a harmonic trapping potential, the problem of two atoms interacting isotropically is separable in centre-of-mass and relative motion. A cylindrically symmetric harmonic trap can be made either a quasi-two dimensional (quasi-2D) or quasi-one dimensional by changing the ratio of the radial to axial size of the trap very large or very small, respectively. Since the radial (axial) size of a harmonic trap is inversely proportional to the square root of the frequency of the radial (axial) harmonic potential, this ratio between the two sizes can be made quite large (small) by changing the ratio of the axial to radial frequency quite large (small). Interacting atomic systems in such quasi-low dimensions are important for studying low dimensional physics. In a seminal paper, Busch et al \cite{Englert:1998} have derived exact analytical solutions of two atoms interacting via zero-range regularized delta potential in an isotropic 3D, 2D and 1D harmonic trap. These solutions clearly show how bound states in harmonic oscillator are modified due to zero-range interactions. For an axially symmetric trap, exact solution of two interacting particles was calculated by Idziaszek and Calarco \cite{Idziaszek_05} under pseudopotential approximation. The validity of the calculation of ref. \cite{Englert:1998} is restricted to the sufficiently weak trap where trap width is much larger than $|a_s|$ \cite{Tiesinga_2000}. To overcome this limitation Bolda and coworkers \cite{Bolda_02} have introduced energy dependent scattering length in regularized pseudopotential and developed a self-consistent method to study two-atom bound states in isotropic traps and $s$-wave collisional properties in optical lattice with axially symmetric quasi-one and two dimensional harmonic confinement \cite{Bolda_03}. 2D and 1D harmonic traps are two ideal cases, and in real situation these lower dimensionality can be realized by considering tight confinement in one or two spatial directions of a 3D trap. Petrov and Shlyapnikov have shown quasi-2D and 3D regimes of atom-atom scattering of a tight axially confined trap \cite{Petrov_01}. Effects of this tight confinement on Bose-Einstein condensates (BEC) is an interesting topic of research both theoretically \cite{You_96,Ketterle_96,Ketterle_97,Petrov_2000,Dalibard_07,Uncu_07,Armijo_11} and experimentally \cite{Gorlitz_01,Schreck_01,Kinoshita,RuGway_13}. For tightly confined atomic gases, the size of the trap may become comparable to the scattering length or the range of interaction. In such situations, it is not clear how the trapping states will be influenced by the range of interactions.

Here we investigate into the bound state properties of a pair of particles interacting with a finite-range model potential in isotropic and quasi-1D harmonic traps. The purpose of our investigation is to understand the effects of finite-range on two-body bound states in isotropic and low dimensional traps. We have used two model potentials - one for positive and the other for negative scattering length. These two potentials do not diverge as $a_{s} \rightarrow \pm \infty$ unlike delta function potential. In the limits $a_{s} \rightarrow \pm \infty$, both potentials reduce to the same form. The model potentials we consider are two-parameter potentials that are shown to be useful for describing the effects of finite-range of interaction near the unitarity regime \cite{Ijmp:Deb_2016}. The two parameters are the range ($r_{0}$) of the potential and the $s$-wave scattering length $a_{s}$. The finite-range potential for negative scattering length is of the form
\begin{equation}
 V_{-}(r) = - \frac{4 \hbar^2}{ \mu r_0^2}  \frac { \alpha \beta^2 }{ [ \alpha \exp(\beta r/r_0) + \exp(-\beta r/r_0) ]^2} 
\label{negative}
\end{equation}
where $\alpha = \sqrt{1 - 2 r_0/a_s}$,  $\beta =  1 + \alpha $ and $\mu$ is the reduced mass. 
For positive scattering length, the potential has the form
\begin{equation}
 V_{+}(r)=   - \frac{4 \hbar^2}{ \mu r_0^2}  \frac { \alpha \beta^2 }{ [ \exp(\beta r/r_0) + \alpha \exp(- \beta r/r_0) ]^2}.  
\label{positive}
\end{equation}
These potentials are derived based on finite-range expansion of $s$-wave phase shift $\delta_0(E)$, where $E$ is the collision energy of two particles colliding in free space at low energy. The potential of Eq. (1) is valid for $|\delta_0(E)| < \pi/2$ in the limit $E \rightarrow 0$. For $|a_s|\gg r_0$, both the potentials of Eqs. (1) and (2) reduce to the form 
\begin{equation}
V_{\pm}(r) = V_{0}(r) + V_{\pm}^{(\epsilon)}(r)
\end{equation}
\begin{figure}
\begin{center}
 \includegraphics[width=\columnwidth]{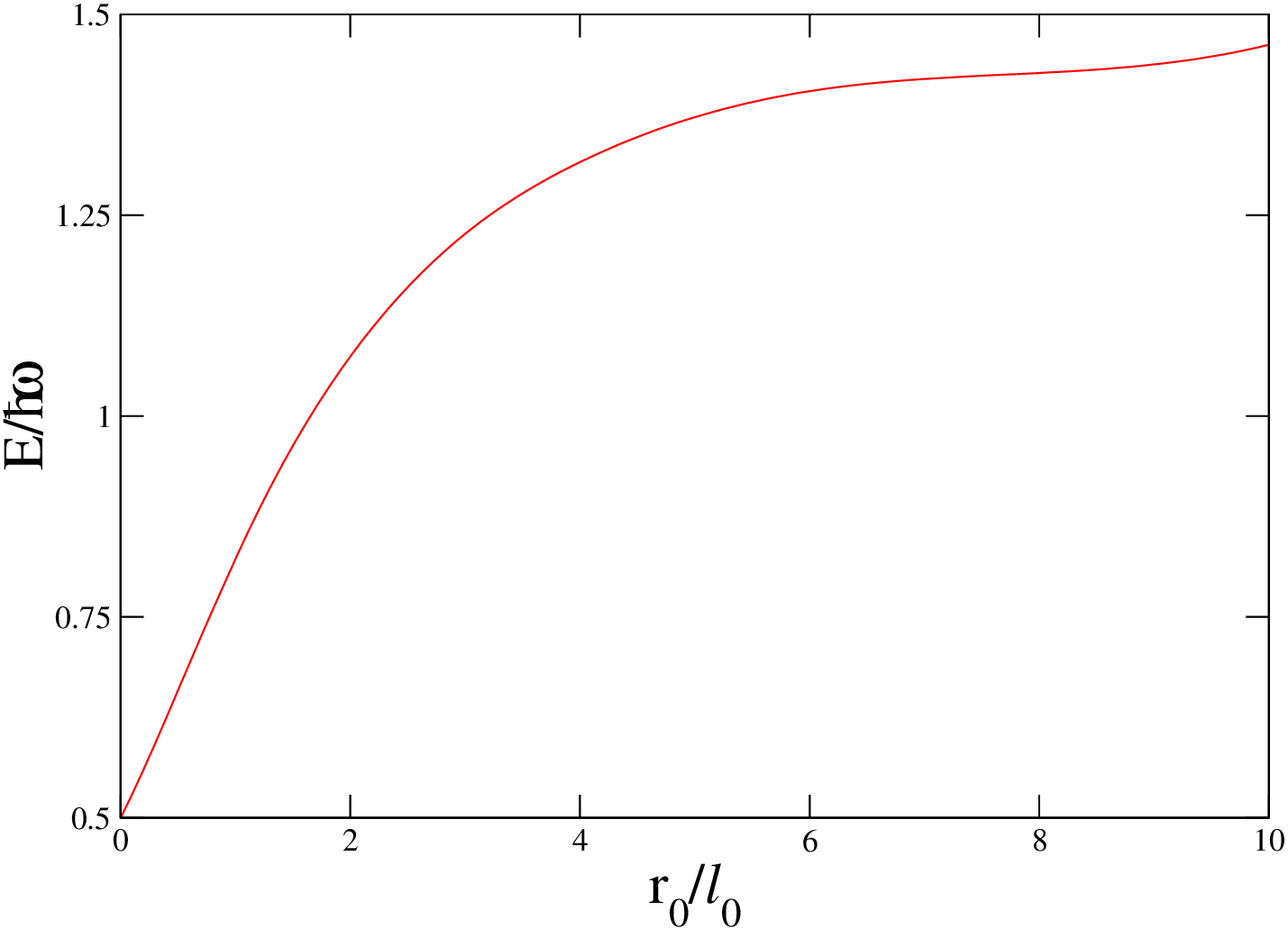}
  \caption{The effects of the range $r_0$ on the ground state energy for two particles interacting via $V_{\infty}$ in an isotropic harmonic trap. For $r_0 \rightarrow 0$, the ground state energy goes to $\frac{1}{2} \hbar\omega$ where $\omega$ is the trapping frequency. Here $l_0 = \sqrt{\frac{\hbar}{\mu\omega}}$ is the harmonic oscillator length scale.}
  \label{fig:1}
\end{center}
\end{figure}
where
\begin{equation} 
V_{0}(r) = -\frac{\hbar\kappa^2}{\mu} \frac{\alpha^{-1}}{\cosh^2[\kappa r]}
\end{equation}
is of the form of well-known P\"{o}schl-Teller potential which is exactly solvable for 1D \cite{Morse,Alhassid_83,Kleinert_92,book:Flugge} and 3D homogeneous systems \cite{Barut1,Barut2}. Here 
\begin{equation} 
V_{\mp}^{(\epsilon)} = V_{0} + \sum_{n = 1}^{\infty} (-1)^{n}(n+1) \left[\frac{\epsilon}{1+\exp(\pm 2\kappa r)}\right]^n
\end{equation}
is proportional to the small parameter $\epsilon = \alpha^{-1} - 1$. $\kappa$ is defined as $\beta/r_0$. The exact solution of $V_{0}$ allows one to write down the Green function of this potential as shown in \cite{Ijmp:Deb_2016}. The potentials of Eq. (\ref{negative}) and (\ref{positive}) are derived based on the effective range expansion of phase shift \cite{Kohn,Kohn2} and in the unitarity limit i.e $a_{s} \rightarrow \pm \infty$ both the potentials reduce to the same form
\begin{equation}
V_{\infty} =   -\frac{4\hbar^2}{\mu r_0^2 \cosh^2(2r/r_0)}
\label{infty}
\end{equation}
This is well known potential and previously used by Shea {\it et al.} \cite{Ajp:Bhaduri:2009} to study energy spectrum of two particles in an isotropic harmonic trap. Here we study the effects of variation of $r_0$ for entire range of scattering length on the bound state properties of two cold atoms in an isotropic 3D harmonic trap as well as tightly confined quasi-1D and 1D trap. Our purpose is to understand the effects of confinement and effective range on the bound states of resonantly interacting cold atoms in lower dimension. Previously, we have used these potentials to investigate bound state properties of two atoms in quasi-two dimension \cite{IJP:Deb:2015}.

\section{Bound states in a harmonic trap}

For two atoms in a harmonic trap, the centre-of-mass and the relative motion between the two atoms become separable. Since an isotropic interaction depends only on the relative coordinate between the two atoms, it will affect the relative motion only. A highly anisotropic harmonic trap is effectively squeezed in one or two spatial directions. It is therefore interesting to know how tight trapping and the low dimensionality affect the bound states of two atoms interacting via a finite-range potential.  
\begin{figure}
 \centering
 \includegraphics[width=\columnwidth]{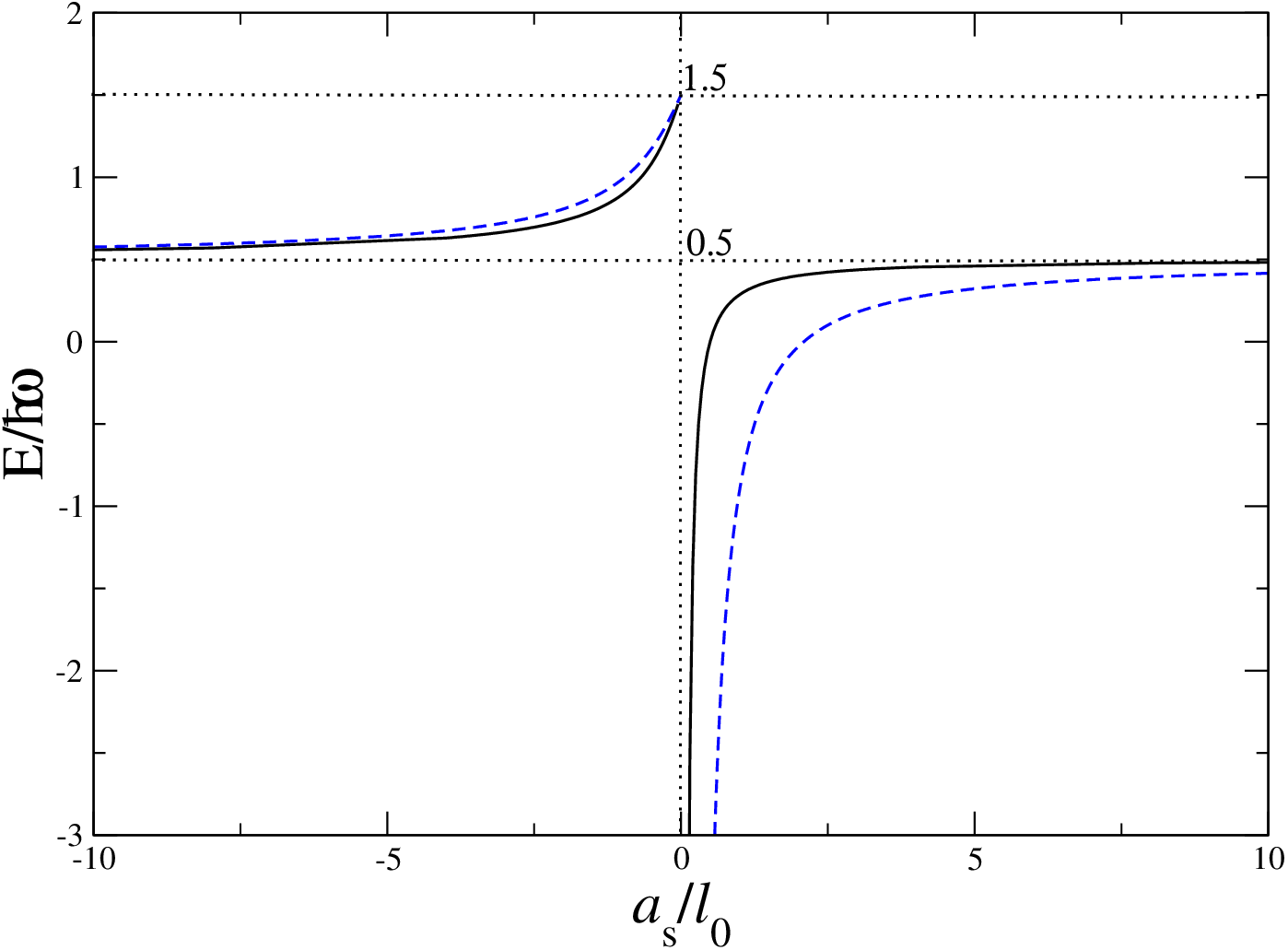}
 \caption{The effects of $a_s$ on ground state energy for two interacting particles in an isotropic harmonic trap considering delta potential (dashed blue line) in one case and the finite-range interaction $V_{\pm}$ (solid black line) in another case, for $r_0 = 0.01 l_0$.}

 \label{fig:3}
\end{figure}

Let us consider an axially symmetric harmonic trap with frequencies $\omega_x = \omega_y = \omega_{\rho} \ne \omega_z$. The shape of an axially symmetric trap depends on the aspect ratio $\eta = \omega_{\rho}/\omega_z$. If $\omega_{\rho}$ $>>$ $\omega_z$ i.e. if the trapping 
frequency $\omega_{\rho}$ in radial direction is much greater than that in axial direction, the trap becomes cigar-shaped (quasi-1D).
For $\omega_z
>> \omega_{\rho} $, the trap has pancake shape (quasi-2D).

\begin{figure}
 \centering
 \includegraphics[width=\columnwidth]{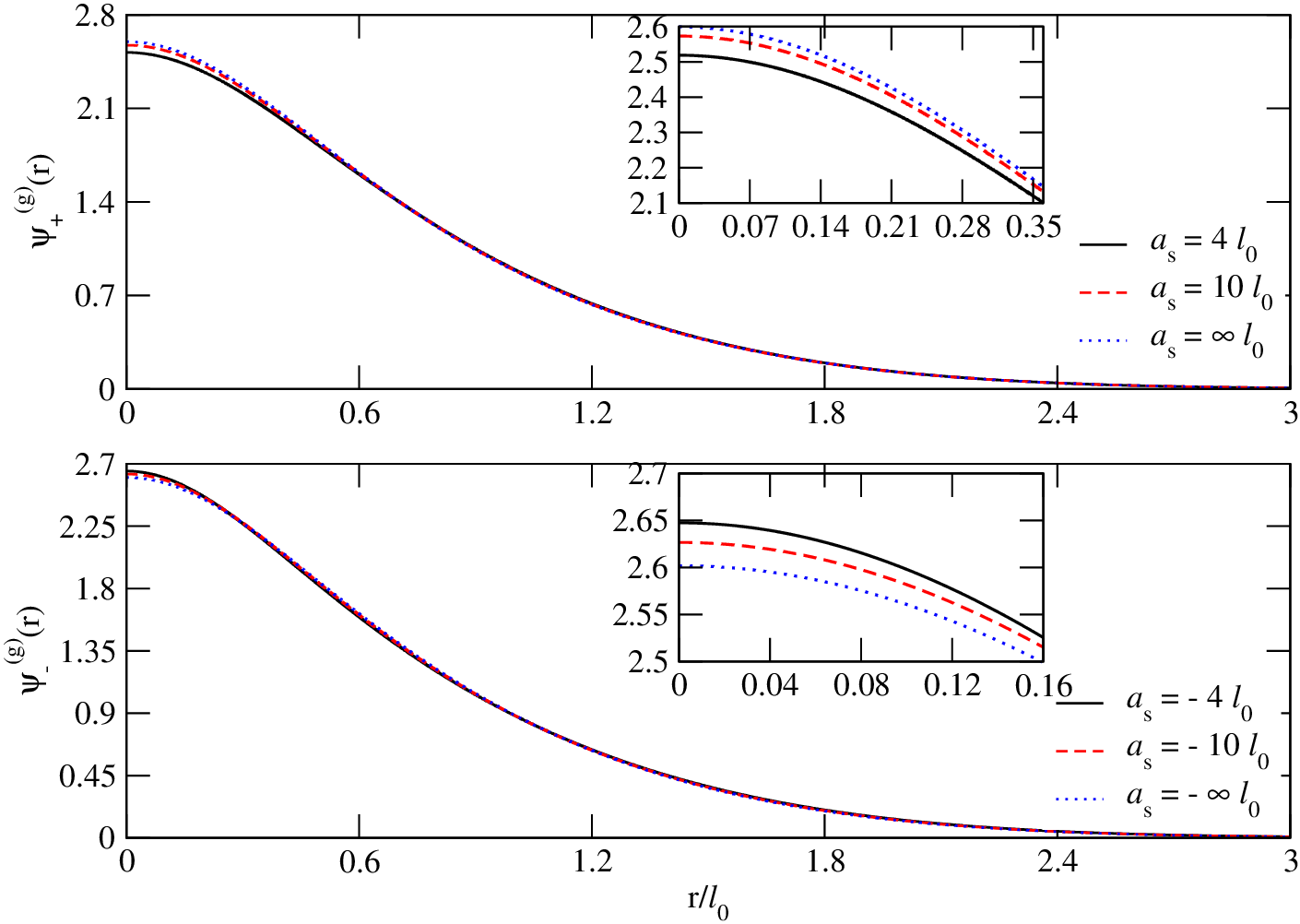}
 \caption{In the upper panel $\Psi_{+}^{(g)}(r)$ in unit of $l_0^{-3/2}$ is plotted for $a_s = 4 l_0$ (solid black line), $a_s = 10 l_0$ (dashed red lines) and $a_s = \infty$ (dotted blue lines) whereas in the lower panel $\Psi_{-}^{(g)}(r)$ in same unit is plotted for $a_s = - 4 l_0$ (solid black line), $a_s = - 10 l_0$ (dashed red lines) and $a_s = - \infty$ (dotted blue lines). In both cases $r_0$ is considered as $1 l_0$.}
\label{fig:4}
\end{figure}

\subsection{3D isotropic harmonic trap}
For two particles in an isotropic harmonic trap, Schr\"{o}dinger equation for relative motion can be written as 
\bea
{}&& \left [-\frac{\hbar^2}{2\mu} \left (\frac{d^2 }{dr^2} -\frac{l(l+1)}{r^2} \right ) + 
\frac{\mu}{2} \omega^2r^2  + V_{s}(r) \right ]u_{s}(r) 
{} =  E u_{s} (r)
\eea
where the subscript $s$ stands for `+' or `-', $\mu$ is the relative mass and $l$ is the azimuthal quantum number. $u_{s} (r)$ is related to 3D wave function $\Psi_{s} (r)$ by $u_{s} (r) = r \Psi_{s} (r)$. Here $E$ denotes the energy eigenvalue of the two interacting atoms. The characteristic length scale of an isotropic trap is defined as $l_0 = \sqrt{\frac{\hbar}{\mu\omega}}$ where $\omega$ is the trapping frequency. 
We solve this radial Schr\"{o}dinger equation by using Numerov method to obtain bound-state wave functions and eigen energies for a wide range of scattering lengths with different effective ranges ($r_0$). For delta potential, the energy eigenvalues are given by the exact solutions of Ref \cite{Englert:1998}
\begin{equation}
 \sqrt{2}\frac{\Gamma(3/4-E/2)}{\Gamma(-E/2+1/4)}=1/a_s
\end{equation}

\subsection{Quasi-1D and 1D}
In case of two interacting atoms in a symmetric harmonic trap, Schr\"{o}dinger equation for relative motion in cylindrical coordinates can be written as
\bea
&& \left [-\frac{\hbar^2}{2\mu} \left (\frac{d^2 }{d\rho^2} +\frac{d}{\rho d\rho} -\frac{|m|^2}{\rho^2} +\frac{d^2}{dz^2} \right ) + 
\frac{\mu}{2} \left (\omega_{\rho}^2\rho^2
   + \omega_z^2 z^2 \right) \right. \nonumber \\ 
 && +  \left. V_{s}(\rho,z) \right ]\Psi_{s}^{a}(\rho,z) =  E_{a} \Psi_{s}^{a} (\rho,z)
\eea
\begin{table}
\caption{First three energy levels in unit of $\hbar \omega_z$ of two particles interacting via $V_{\infty}$ in one dimensional confinement for different values of $r_{0}$.}
\begin{center}
 \begin{tabular}{l c c c}  \hline 
  $r_{0}$    & $E_{0}$      &$E_1$       & $E_2$   \\ \hline 
  10        &0.4605       &1.4622      & 2.4635   \\ 
  5       &0.3510      &1.3702      & 2.3850    \\ 
  1      &-1.9058     &0.8220      & 1.9925     \\ 
  0.1     &-199.9989     &0.5291      & 1.5571      \\ \hline
\end{tabular}
\end{center}
\end{table}
where $E_{a}$ is the total energy and $\Psi_{s}^{a} (\rho,z)$ is the total wave function of the anisotropic trap. Radial and axial modes are coupled via the interaction potential $V_{s}$. Now we consider $\Psi_{n_{\rho},m,{n_z}}^{(a)0}(\rho,z)$ as the wave function of the relative motion between  a pair of noninteracting particles
in the trap, where  $n_{\rho}$ and $n_z$ are the radial and axial principal quantum numbers; and $m$ stands for angular momentum in 2D. This wave function 
is separable in axial and radial coordinates as 
 \begin{equation}
\Psi_{n_{\rho},m,{n_z}}^{(a)0}(\rho,z) = R_{n_{\rho},m}^{0}(\rho) \times f_{n_z}^{0}(z) 
\end{equation}
where
\begin{figure}[!h]
 \centering
 \includegraphics[width=\columnwidth]{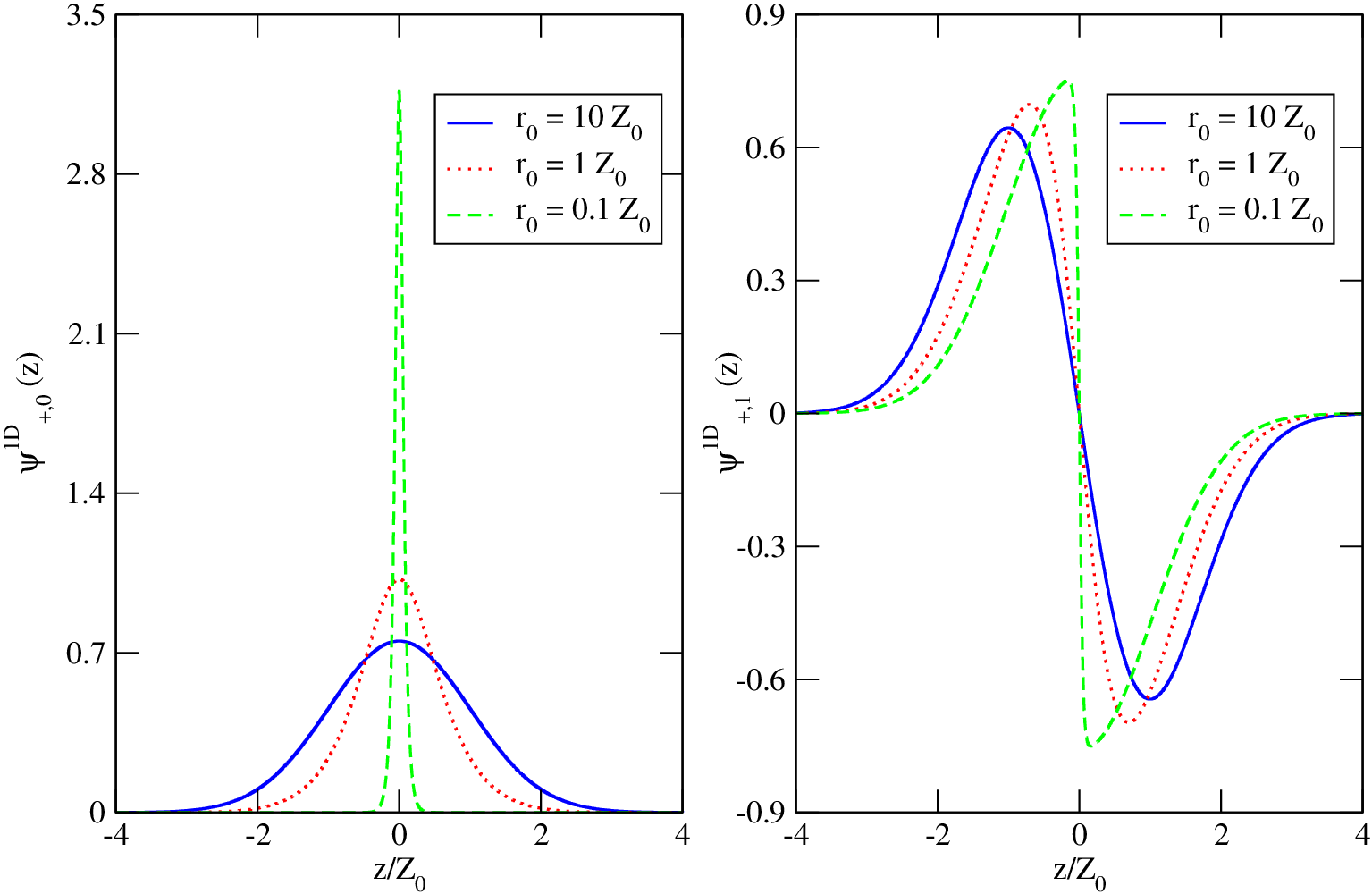}
 \caption{The ground state (left panel) and first excited state (right panel) wave functions in unit of $Z_0^{-1/2}$ for two atoms confined in 1D and interacting via $V_{\infty}$ for $r_0 = 10 Z_0$ (solid blue line), $r_0 = 1 Z_0$ (dotted red lines) and $r_0 = 0.1 Z_0 $ (dashed green lines).}
\label{fig:8}
\end{figure}
\bea
&& R_{n_{\rho},m}^{0}(\rho) = \left[\frac{n_{\rho}!}{\pi\Gamma(n_{\rho}+|m|+1)}\right]^{\frac{1}{2}}\frac{1}{\rho_{0}^{|m|+1}}\rho^{|m|}  \nonumber \\
&& \exp\left[{-\frac{\rho^2}{2 \rho_{0}^2}}\right]  L_{n_{\rho}}^{|m|}\left(\frac{\rho^2}{\rho_{0}^2}\right) 
\eea
and 
\begin{equation}
f_{n_z}^{0}(z) = \frac{\pi^{-\frac{1}{4}}}{\sqrt{2^{n_{z}}{n_z}!}}H_{n_z}\exp\left [-\frac{z^2}{2 Z_{0}^2} \right ]
\end{equation}
$R_{n_{\rho},m}^{0}(\rho)$ and $f_{n_z}^{0}(z)$ are 2D and 1D harmonic oscillator wave functions, respectively. Here 
$L_{n_{\rho}}^{|m|}$ and $H_{n_z}$ are Laguerre and Hermite polynomials, respectively; 
$\rho_{0} = \sqrt{\frac{\hbar}{\mu\omega_{\rho}}}$ and $Z_0 = \sqrt{\frac{\hbar}{\mu\omega_{z}}}$ are the 2D and 1D harmonic oscillator length scales, respectively. These two length scales are related by $Z_0 = \sqrt{\eta} \rho_0$, where $\eta = \omega_{\rho} /\omega_z$ is the aspect ratio.  
For noninteracting case ($V_{s}(\rho,z) = 0$), total energy $E_{a}$ is the sum of 2D and 1D harmonic oscillator eigen energies ($E_{a} = (2n_{\rho}+|m|+1)\hbar\omega_{\rho} +(n_z+1/2)\hbar\omega_z$).
\begin{figure}
 \centering
 \includegraphics[width=\columnwidth]{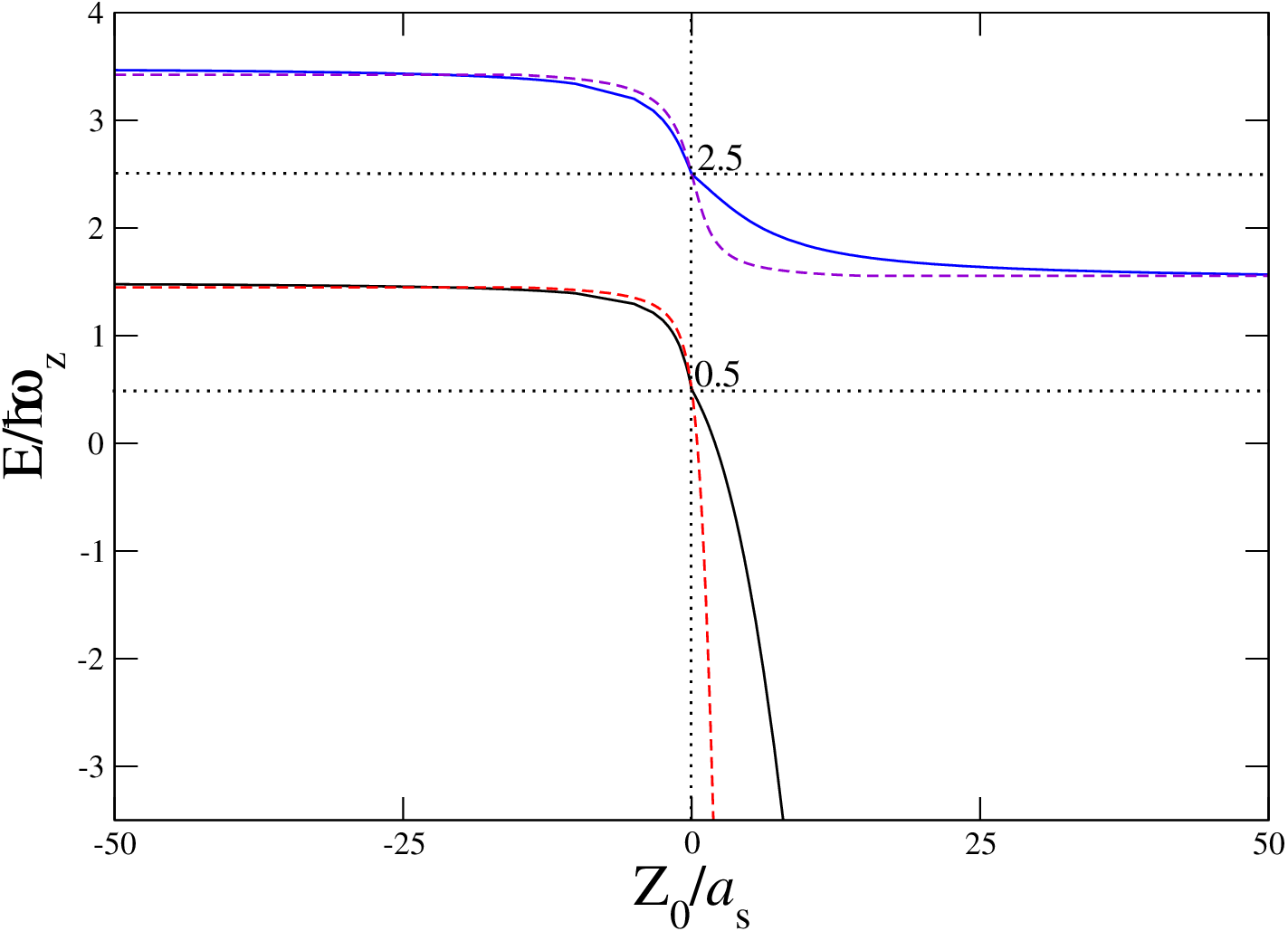}
 \caption{Eigen energies of first two odd state wave functions of two particles in 1D harmonic trap for finite-range interaction (solid line) are plotted as a function of $1/a_s$ in unit of $Z_{0}^{-1}$ with $r_0 = 0.001 Z_0$. Also plotted are the eigen energies of first two even state wave functions of zero-range contact potential (dashed line).}
 \label{fig:10}
\end{figure}

For two atoms confined in one dimension and interacting via delta potential, energy eigenvalue can be determined again from equation (8). Both in 3D and 1D the effective range expansion formula \cite{Adhikari:2000} has the same mathematical form
\begin{equation}
k \cot \delta = -\frac{1}{a_s} + \frac{1}{2} r_0 k^2 + ...
\end{equation}
Since our model potentials are derived by the method of Jost and Kohn \cite{Kohn}, based on effective range expansion, the model potentials have the same form both in 1D and 3D. In case of trapped atoms, one can achieve quasi-1D situation while exactly one dimensional atomic gas is an ideal system. 
\begin{figure}
\centering
 \includegraphics[width=\columnwidth]{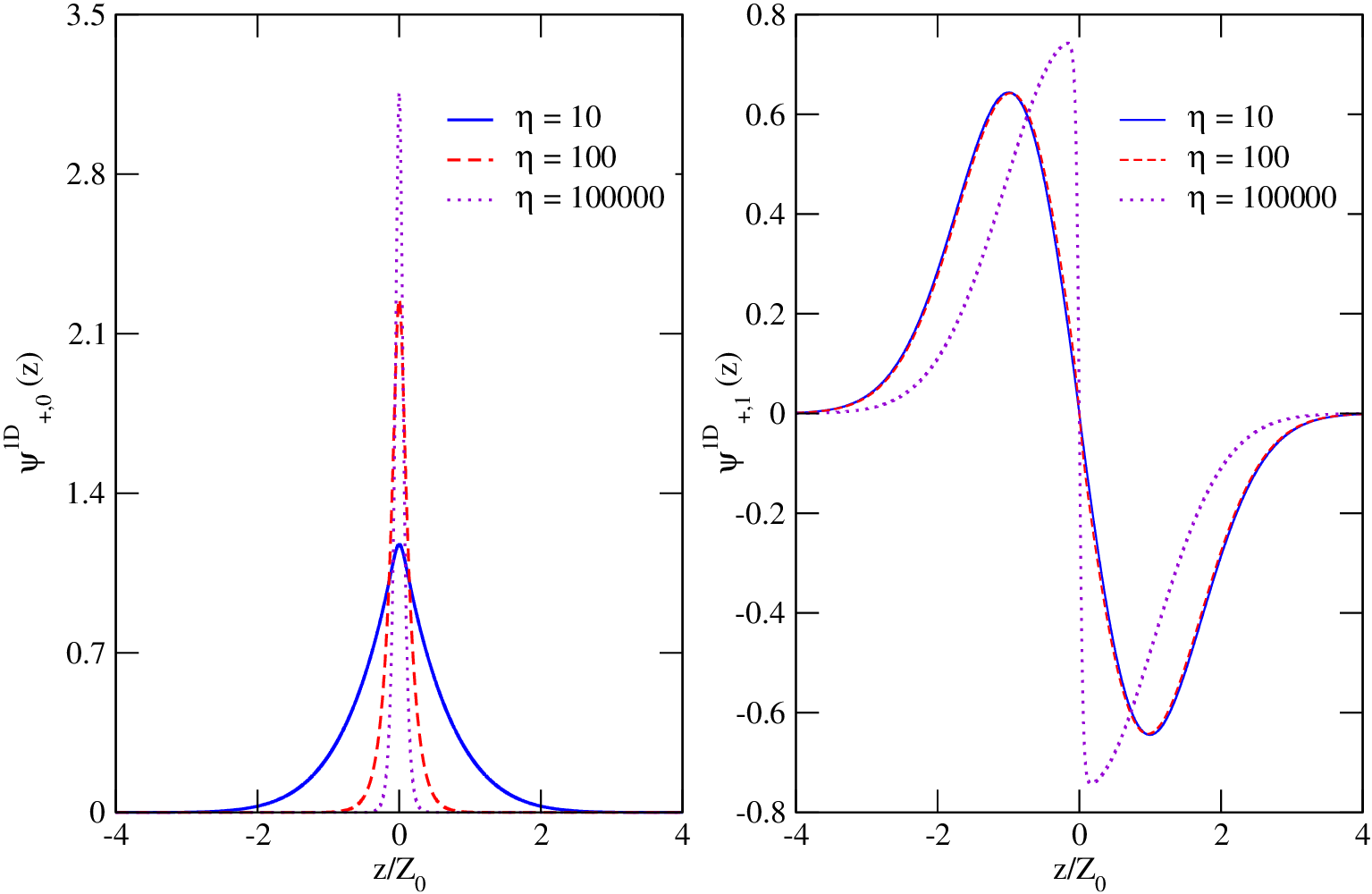}
 \caption{Ground and first excited state wave functions (in unit of $Z_{0}^{-1/2}$) of two atoms interacting via $V_{\infty}$ are plotted in left and right panels respectively for $\eta = 10$ (solid blue line), $\eta = 100$ (dashed red lines) and $\eta = 100000$ (dotted violet lines). The value of $r_0$ is $0.1 Z_0$. } \label{fig:11}
\end{figure}

Now, a quantum system is effectively one dimensional if the chemical potential and thermal energies are smaller than energy gap in strongly confined direction i.e radial direction. When the frequency in radial direction is much greater than axial frequency, spread of the wave function  is narrow in radial direction but much widely spaced along the axial direction. The total wave function is then very much closer to that of an one dimensional harmonic oscillator. Under such conditions, the radial wave function is effectively the ground state wave function of simple 2D harmonic oscillator. Then total wave function can be written as $\Psi_{s}^{a}(\rho,z)$=$R_{0,0}^0(\rho)\Psi_{s,n}^{1D}(z)$. $\Psi_{s,n}^{1D}$ is the reduced wave function in the axial direction. $n$ carries the same significance as $n_z$ and equal to the number of nodes in the wave function. Depending upon whether the number of nodal points are even or odd, the wave functions are designated as even or odd state, respectively. Projecting both sides of equation (9) with $R_{0,0}^0(\rho)$, Schr\"{o}dinger equation reduces to the form

\begin{equation}
  \left [-\frac{\hbar^2}{2\mu} \left (\frac{d^2}{dz^2} \right ) + 
 \frac{\mu}{2} \omega_z^2 z^2 + V_{s}(z) \right ]\Psi_{s,n}^{1D}(z) =  E_{n}^{1D} \Psi_{s,n}^{1D} (z)
\end{equation}
  where $V_{s}(z)$ is the effective interaction in 1D given by
\begin{equation}
 V_{s}(z) = \int 2 \pi (R_{0,0}^0)^2 V_{s}(\rho,z) \rho d\rho 
\end{equation} 
$E_{a} - \hbar\omega_{\rho} = E_{n}^{1D}$ is the reduced energy. We have used $E_{n_z}$ to represent the pure 1D energy. 
\begin{table}
\caption{First three energy levels in unit of $\hbar \omega_z$ of two particles interacting via $V_{\infty}$ in quasi-one dimensional confinement for different values of $r_0$ and $\eta$. }
\begin{center}
 \begin{tabular}{lcccc}  \hline 
  $r_0$    & $\eta$    & $E_{0}^{1D}$      &$E_{1}^{1D}$       & $E_{2}^{1D}$   \\ \hline
         &10      &0.4605      & 1.4622    & 2.4635  \\ 
  10       &100      &0.4605      & 1.4622      & 2.4635   \\ 
        &1000     &0.4605      & 1.4622       & 2.4635     \\ \hline
       & 10             &0.3532      & 1.3722  & 2.3865    \\ 
  5    &  100            &0.3512      & 1.3705   & 2.3852  \\  
     &   1000           &0.3510      & 1.3702   & 2.3850    \\  \hline
    &   10             & -1.2285    & 1.0021   & 2.0935     \\  
  1   &   100            & -1.8097    & 0.8472   & 2.006      \\   
     &   1000           & -1.8957    & 0.8245   & 1.9941     \\   \hline  
  &10        &-0.9315       &1.4945      & 2.1205   \\ 
  &100       &-28.9427      &1.4570      & 1.6645    \\ 
0.1 &  1000      &-134.9057     &1.2245      & 1.5760     \\ 
  &10000     &-190.6745     &0.7131      & 1.5595      \\ 
  &100000    &-199.0217              &0.5505      & 1.5572      \\  \hline 
\end{tabular}
\end{center}
\end{table}

\begin{table}
\caption{First three energy levels in unit of $\hbar \omega_z$ of two particles interacting via $V_{\pm}$ in quasi-one dimensional confinement for $\eta =100000$ with $r_0 = 0.1 Z_0$. }
\begin{center}
 \begin{tabular}{lccc}  \hline 
$a_s$    & $E_{0}^{1D}$      &$E_{1}^{1D}$       & $E_{2}^{1D}$  \\ \hline
0.2      & -161.395    & -3.8938    & 1.4892  \\ 
0.5      & -191.087     & 0.0032      & 1.5535  \\  
1        & -195.342     & 0.3275         &  1.5572       \\    
10       & -198.674     & 0.5505         &  1.5572        \\    
-10      & -199.975     & 0.6041         &  1.5572        \\     
-1       & -207.945     & 0.9281         &  1.5572        \\     
-0.5     & -215.736     & 1.1060          &  1.5571        \\     
-0.2     & -234.554     & 1.3035         &  1.5562        \\      \hline 
\end{tabular}
\end{center}
\end{table}

\subsection{Results and discussions}
To check whether one can retrieve the standard results of zero-range interactions, we first discuss the results in the limit $r_0 \rightarrow 0$. 
  For two interacting atoms in an isotropic trap, lowering the range decreases the energy level. As $r_0 \rightarrow 0$ energy level shifts towards lower energy. This is shown in Fig. 1. In Fig. 2 we have shown how the scattering length affects the ground state energy. In the limit $a_s \rightarrow 0^{-}$, the ground state of relative motion in the trap reduces to that of a noninteracting pair of particles. In free space, there exists only one bound state when $a_s > 0$. In the limit $a_s \rightarrow 0^{+}$, the binding energy of the bound state in free space diverges as $1/a_{s}^2$. This fact is also reflected in the case of isotropic trap as the ground state energy diverges to the large negative values in the limit $a_s \rightarrow 0^{+}$. In the limit $r_0 \rightarrow 0$, these results are similar to the results already discussed in \cite{Englert:1998, Ajp:Bhaduri:2009}. In Fig. 3 we have shown the effects of $a_s$ on ground and first excited state wave functions. Increase in scattering length increases the probability of finding the particle near the trap centre.

Now, we consider two particles in 1D confinement. We have used the same form of interaction for 1D harmonic trap, with the scattering length being replaced by that of 1D. In table 1 we have shown how eigen energies depend on $r_0$. The ground state energy diverges as we move from high $r_0$ to low $r_0$. In table 2 we have shown the effects of $\eta$ and $r_0$ on first three energy levels for $V_{\infty}$ in quasi-one dimensional trap. Now, if we compare these results with those in table 1 we can conclude that for large range small aspect ratio ($\eta$) is reasonably good to produce purely 1D results. For $r_0 = 0.1 Z_0$, quasi-1D to 1D transition occurs at $\eta = 100000$.

In Fig. 4 we have plotted ground and first excited wave functions for three different $r_0$. It is evident from the figure that lowering $r_0$ squeezes the wave function to the centre. The variation of the eigen energies of odd states with $a_s$ for low range interaction is similar to that of the variation of even state of same energy for zero range interaction as shown in Fig. 5. In Fig. 6 we have shown the effects of $\eta$ on ground and first excited states for quasi-1D trap. As $\eta$ increases wave function squeezes more towards the trap centre. For finite-range potential in quasi-1D, the even states are weakly dependent on $a_s$ whereas the odd states change significantly as shown in table 3.

\section{Conclusions}
In conclusion, we have presented a model study showing the effects of finite-range $r_0$ of interaction 
on the bound state properties of two particles in isotropic 3D, quasi-1D and 1D traps.
We have shown that, in case of isotropic 3D trap, the results of a contact potential \cite{Englert:1998} 
can be reproduced with our model potentials by taking the limit $r_0 \rightarrow 0$. We have examined whether a finite-range 1D model potential can
reproduce the results of 1D contact potential when the range goes to zero. In contrast to 3D case, we find that the results for finite-range 1D
potentials do not approach to those for 1D contact potential as the range decreases. 
This indicates that probably one can not realize the theoretical results  \cite{Englert:1998} of two particles interacting with 1D contact 
potential in a 1D harmonic trap. Remarkably, the 1D finite-range model potentials
have the same mathematical form as the 3D ones, because the finite-range expansion of phase shift  both in 3D
and 1D is similar. We find that in the limit of large range,  the bound state energies
in  1D or 3D reduce to those of non-interacting two particles in 1D or 3D trap, respectively. In the case of quasi-1D or 1D, as the range decreases below
the 1D harmonic oscillator length scale, the ground state energy decreases below zero. We have demonstrated how quasi-1D results approach 
to 1D results as the aspect ratio $\eta$ is increased. We find that for smaller range, 1D results are nearly reproduced when $\eta$ is larger 
than 10000. However, if the range is large, one can obtain nearly 1D results with smaller $\eta$ ($10000> \eta > 10$). We have studied 
finite-range interactions over the entire range of scattering length including the unitarity regime $a_s \rightarrow \pm \infty$. 
This study will be important for exploring physics of trapped one or quasi-one dimensional atoms interacting with finite 
effective range. For instance, the effective range effects in ultracold atoms interacting with a narrow Feshbach resonance 
can not be neglected. 

\vspace{0.5cm} 

\noindent 
{\bf Acknowledgements} \\
Partha Goswami is thankful to CSIR, Govt. of India, for support.
 
\section*{References}
\bibliography{int_trap2}

\providecommand{\newblock}{}
\begin{thebibliography}{10}
\expandafter\ifx\csname url\endcsname\relax
  \def\url#1{{\tt #1}}\fi
\expandafter\ifx\csname urlprefix\endcsname\relax\def\urlprefix{URL }\fi
\providecommand{\eprint}[2][]{\url{#2}}

\bibitem{Tonks}
Tonks L 1936 {\em Phys. Rev.\/} {\bf 50} 955

\bibitem{Girardeau}
Girardeau M 1960 {\em J. Math. Phys.\/} {\bf 1} 516

\bibitem{Lieb}
Lieb E~H and Liniger W 1963 {\em Phys. Rev.\/} {\bf 130} 1605

\bibitem{Paredes}
Paredes B {\em et~al.\/} 2004 {\em Nature\/} {\bf 429} 277

\bibitem{Kinoshita}
Kinoshita T, Wenger T and Weiss D~S 2004 {\em Science\/} {\bf 305} 1125

\bibitem{PRL:Olshanii:1998}
Olshanii M 1998 {\em Phys. Rev. Lett.\/} {\bf 81} 5

\bibitem{PRL:Olshanii:2003}
Bergman T, Moore M~G and Olshanii M 2003 {\em Phys. Rev. Lett.\/} {\bf 91} 16

\bibitem{PRL:Moritz:2005}
Moritz H {\em et~al.\/} 2005 {\em Phys. Rev. Lett.\/} {\bf 94} 210401

\bibitem{Kagan_2000}
Kagan Y, Prokof'ev N~V and Svistunov B~V 2000 {\em Phys. Rev. A\/} {\bf 61}
  045601

\bibitem{Petrov_01}
Petrov D~S and Shlyapnikov G~V 2001 {\em Phys. Rev. A\/} {\bf 64} 012706

\bibitem{Bolda_03}
Bolda E~L, Tiesinga E and Julienne P~S 2003 {\em Phys. Rev. A\/} {\bf 68}
  032702

\bibitem{Idziaszek_05}
Idziaszek Z and Calarco T 2005 {\em Phys. Rev. A\/} {\bf 71} 050701(R)

\bibitem{Chin_11}
Hung C~L, Zhang X, Gemelke N and Chin C 2011 {\em Nature\/} {\bf 470} 236

\bibitem{Chevy_02}
Chevy F, Bretin V, Rosenbusch P, Madison K~W and Dalibard J 2002 {\em Phys.
  Rev. Lett.\/} {\bf 88} 250402

\bibitem{Vogt_12}
Vogt E {\em et~al.\/} 2012 {\em Phys. Rev. Lett.\/} {\bf 108} 070404

\bibitem{Perin_13}
Merloti K, Dubessy R, Longchambon L, MOlshanii and Perrin H 2013 {\em Phys.
  Rev. A\/} {\bf 88} 061603

\bibitem{Baskaran_89}
Baskaran G 1989 {\em Phys. Scr.\/} {\bf T27} 53

\bibitem{Englert:1998}
Busch T, Englert B~G, Rzazewski K and Wilkens M 1998 {\em Foundation of
  Physics\/} {\bf 28} 549

\bibitem{IJP:Deb:2015}
Goswami P, Rakshit A and Deb B 2015 {\em Indian J. Phys.\/} {\bf 89} 8

\bibitem{Ijmp:Deb_2016}
Deb B 2016 {\em Int. J. Mod. Phys. B\/} {\bf 30} 1650036

\bibitem{Stoof_93}
Tiesinga E, Verhaar B~J and Stoof H~T~C 1993 {\em Phys. Rev. A\/} {\bf 47} 4114

\bibitem{Kohler_06}
K{\"o}hler T, G{\'o}ral K and Julienne P~S 2006 {\em Rev. Mod. Phys.\/} {\bf
  78} 1311

\bibitem{Chin_10}
Chin C, Grimm R, Julienne P and Tiesinga E 2010 {\em Rev. Mod. Phys.\/} {\bf
  82} 1225

\bibitem{Harabati_99}
Flambaun V~V, Gribakin G~F and Harabati C 1999 {\em Phys. Rev. A\/} {\bf 59} 3

\bibitem{Stoof_04}
Duine R~A and Stoof H~T~C 2004 {\em Physics Reports\/} {\bf 396} 115

\bibitem{Pethick_05}
Schwenk A and Pethick C~J 2005 {\em Phys. Rev. Lett.\/} {\bf 95} 160401

\bibitem{Castin_06}
Massignan P and Castin Y 2006 {\em Phys. Rev. A\/} {\bf 74} 013616

\bibitem{Zoller_08}
Pupillo G, Griessner A, Micheli A, Ortner M, Wang D~W and Zoller P 2008 {\em
  Phys. Rev. Lett.\/} {\bf 100} 050402

\bibitem{Li_12}
Ho T~L, Cui X and Li W 2012 {\em Phys. Rev. Lett.\/} {\bf 108} 250401

\bibitem{Ohara_12}
Hazlett E~L, Zhang Y, Stites R~W and \'{O}Hara K~M 2012 {\em Phys. Rev.
  Lett.\/} {\bf 108} 045304

\bibitem{Thomas_12}
Wu H and Thomas J~E 2012 {\em Phys. Rev. A\/} {\bf 86} 063625

\bibitem{Tiesinga_2000}
Tiesinga E, Williams C~J, Mies F~H and Julienne P~S 2000 {\em Phys. Rev. A\/}
  {\bf 61} 063416

\bibitem{Bolda_02}
Bolda E~L, Tiesinga E and Julienne P~S 2002 {\em Phys. Rev. A\/} {\bf 61}
  013403

\bibitem{You_96}
Lewenstein M and You L 1996 {\em Phys. Rev. A\/} {\bf 53} 909

\bibitem{Ketterle_96}
Ketterle W and van Druten N~J 1996 {\em Phys. Rev. A\/} {\bf 54} 656

\bibitem{Ketterle_97}
Ketterle W and van Druten N~J 1997 {\em Phys. Rev. A\/} {\bf 79} 549

\bibitem{Petrov_2000}
Petrov D~S, Shlyapnikov G~V and Walraven J~T~M 2000 {\em Phys. Rev. Lett.\/}
  {\bf 85} 3745

\bibitem{Dalibard_07}
Kruger P, Hadzibabic Z and Dalibard J 2007 {\em Phys. Rev. Lett.\/} {\bf 99}
  040402

\bibitem{Uncu_07}
Uncu H, Tarhan D, Demiralp E and Mustecaplioglu O~E 2007 {\em Phys. Rev. A\/}
  {\bf 76} 013618

\bibitem{Armijo_11}
Armijo J, Jacqmin T, Kheruntsyan K~V and Bouchoule I 2011 {\em Phys. Rev. A\/}
  {\bf 83}(R) 021605

\bibitem{Gorlitz_01}
G{\"o}rlitz A {\em et~al.\/} 2001 {\em Phys. Rev. Lett.\/} {\bf 87} 130402

\bibitem{Schreck_01}
Schreck F {\em et~al.\/} 2001 {\em Phys. Rev. Lett.\/} {\bf 87} 080403

\bibitem{RuGway_13}
RuGway W {\em et~al.\/} 2013 {\em Phys. Rev. Lett.\/} {\bf 111} 093601

\bibitem{Morse}
Rosen N and Morse P~M 1932 {\em Phys. Rev.\/} {\bf 42} 210

\bibitem{Alhassid_83}
Alhassid Y, G{\"u}rsey F and Iachello F 1983 {\em Phys. Rev. Lett.\/} {\bf 50}
  873

\bibitem{Kleinert_92}
Kleinert H and Mustapic I 1992 {\em J. Math. Phys.\/} {\bf 33} 643

\bibitem{book:Flugge}
Fl{\"u}gge S 1994 {\em Practical Quantum Mechanics\/} (Berlin: Springer)

\bibitem{Barut1}
Barut A~O, Inomata A and Wilson R 1987 {\em J. Phys. A: Math. Gen\/} {\bf 20}
  4075

\bibitem{Barut2}
Barut A~O, Inomata A and Wilson R 1987 {\em J. Phys. A: Math. Gen\/} {\bf 20}
  4083

\bibitem{Kohn}
Jost R and Kohn W 1952 {\em Phys. Rev.\/} {\bf 87} 6

\bibitem{Kohn2}
Jost R and Kohn W 1953 {\em Dan. Mat. Fys. Medd.\/} {\bf 27} 9

\bibitem{Ajp:Bhaduri:2009}
Shea P, van Zyl B~P and Bhaduri R~K 2009 {\em Am. J. {P}hys.\/} {\bf 77} 511

\bibitem{Adhikari:2000}
Barlette V~E, Leite M~M and Adhikari S~K 2000 {\em Eur. J. Phys.\/} {\bf 21}
  435

\end{thebibliography}

\end{document}